\documentstyle[11pt,newpasp,twoside]{article}
\def\solm{M$_{\odot}\,$}

\def\edcomment#1{\iffalse\marginpar{\raggedright\sl#1\/}\else\relax\fi}
\reversemarginpar

\begin{document}
\title{The Relationship between Stellar and Halo Masses of Disk Galaxies at $z = 0.2 - 1.2$}
 \author{Christopher J. Conselice$^{1}$, Kevin Bundy$^{1}$, Richard S. Ellis$^{1}$ \\ Jarle Brinchmann$^{2}$, Nicole Vogt$^{3}$}
\affil{(1) California Institute of Technology, (2) Max Planck Institute for Astrophysics, (3) New Mexico State University}

\begin{abstract}

We present the results of a study to determine the co-evolution of
the virial and stellar masses for a sample of 83 disk galaxies between 
redshifts
$z = 0.2 - 1.2$.  The virial masses of these disks are computed
using measured maximum rotational velocities from Keck 
spectroscopy and scale lengths from Hubble Space Telescope imaging.
We compute stellar masses based on stellar population synthesis model
fits to spectral energy distributions including K(2.2$\mu$m) band 
magnitudes.   We find no apparent evolution with
redshift from $z = 0.2 - 1.2$ in the relationship between stellar masses 
and maximum rotational velocities through the
stellar mass Tully-Fisher relationship.  We also find no evolution  when 
comparing disk stellar and virial masses.
Massive disk galaxies therefore appear to be already in place,
in terms of their virial and stellar masses, out to the highest redshifts
where they can be morphologically identified.

\end{abstract}

\section{Introduction}

It is now widely held that most galaxies are  embedded in dark matter halos.   
These halos form from fluctuations 
of largely non-baryonic mass in the early universe and initially also 
contain baryons in the form of a hot
gaseous phase.  This gas later cools, leading to star formation.
The history of this star formation, and the resulting build up of stellar
mass,
is now being traced out (e.g., Madau et al. 1998), yet the physical
details of mass assembly remain a major observational challenge.
The fraction of gas mass converted into stars is regulated by
poorly understood processes such as feedback, cooling, and star formation
(e.g., van den Bosch 2002).  Theoretical methods, such as those in 
semi-analytic simulations (e.g., Abadi et al. 2003), predict when galaxies 
form their 
stellar mass. It is however preferable to directly measure this 
evolution through observations.  Perhaps the most suitable galaxy type
for performing this experiment is disk galaxies where the dark halo and 
galaxy appear intimately tied, and virial and stellar masses are in principle 
directly measurable. 

Early work on this problem has concentrated on the Tully-Fisher (TF)
relationship, which traces the tight correlation between the luminosities 
and rotational velocities
of disk galaxies  (Verheijen 2001).  Out to redshifts of $z \sim 1$ there 
appears to be only a slight evolution in this relationship, such that disks
are at most $\sim$ 0.5 magnitudes brighter at a given rotational velocity in 
comparison to those at $z \sim 1$ (Vogt et al. 1996, 1997; Ziegler 
et al. 2002; Bohm et al. 2003).   The fact that the TF relationship does
not evolve strongly with time may mask some interesting trends, as the 
luminosity and rotational velocities of disks can be affected by more
than just stellar and dynamical masses.  As such, we 
have begun a study to directly determine the relationship and evolution
between the
stellar and virial masses of a sample of disk galaxies out to $z \sim 1.2$.   
We use these data to address several 
issues including: do massive disks form their virial mass all at once, or 
does material accrete onto a disk over time?  Is baryonic
material slowly being converted into stars in disks at $z < 1$, or is most 
stellar mass in place by then? Our major conclusion is that
we do not find significant evolution in the relationship
between stellar and virial masses with redshift.

\section{Data and Analysis}

The sample used in this paper consists of 83 disk galaxies at redshifts
from $z \sim 0.2$ to $z \sim 1.2$, with an average redshift of
$z \sim 0.7$.  These galaxies were taken from the Groth
Strip Field Survey (Vogt et al. 1996, 1997, 2004 in prep). 
These galaxies were typically selected morphologically based on the appearance
of a discernible disk in Hubble Space Telescope (HST) F814W (I$_{814}$) images,
and in all cases with an apparent
magnitude upper limit (I$_{814}$ = 22.5) for which spectroscopy can be 
acquired 
(Vogt et al. 1997).    We also took deep Keck or UKIRT K-band imaging for
each of these systems.
Our sample is therefore composed of several different
data sets which as a group, or individually, are not homogeneous or complete
in any sense.  There are three main data products used in this analysis: HST 
imaging for measuring scale-lengths and optical photometry, K-band
imaging, and spectroscopy for measuring V$_{\rm max}$ values.

The spectroscopy was acquired with Keck LRIS by observing each disk 
along its major axis, which was determined using HST images (see e.g.,
Vogt et al. 1997).  The rotation curve for most systems is
sampled through the [OII] line, which is visible in the observed
optical for galaxies up to $z \sim 1$.   Once these raw rotation curves
are constructed they are fit to a model that assumes some maximum velocity at 
one and a half times the scale length (R$_{\rm d}$) (Vogt et al. 1996).   
At radii larger than 1.5$\times$ R$_{\rm d}$, the velocity curve in this
model remains at the
maximum.  This model rotation curve grows linearly from zero velocity at the 
center of the 
galaxy to V$_{\rm max}$ at 1.5$\times$ R$_{\rm d}$ (Vogt et al. 1996).
The observed profile is then
fit to various forms of these models, taking into account the seeing, slit 
width, and misalignment with the major axis,  until a best fit solution is 
found from which the maximum velocity, V$_{\rm max}$, is retrieved.  We use
these V$_{\rm max}$ values to compute our virial masses (\S 3).

\section{Virial and Stellar Masses}

Simulations have shown that neither the 
K-band light nor the velocity (or velocity squared), are necessarily
good traces of the total
masses of disk galaxies (e.g., van den Bosch 2002).   
Using a series of models, van den Bosch (2002) showed
through simulations that the quantity R$_{\rm d}$V$_{\rm max}$$^2$/G gives,
on average, a good representation of virial mass to within
50\%. The best fitting empirical formula, found by van
den Bosch (2002), is:

\begin{equation}
M_{\rm vir} = 2.54 \times 10^{10} M_{\odot}\  \left(\frac{R_{\rm d}}{{\rm kpc}}\right)\left(\frac{V_{\rm max}}{100\, {\rm km\, s^{-1}}}\right)^{2},
\end{equation}

\noindent where R$_{\rm d}$ is the scale length fit and 
$V_{\rm max}$ is the maximum retrieved velocity of the rotation curves. 
The zero point of the relationship between R$_{\rm d}$V$_{\rm max}$$^2$/G
and virial masses is found to be independent of the amount of 
feedback and pre-enrichment (van den Bosch 2000).   By using this formula we 
are making the assumption that the maximum rotation velocities derived from the
LRIS spectra represent the largest observable velocity of each galaxy. 

Stellar masses are based on analysis of our HST and
K-band photometry. We use this photometry,
in typically three bands, the K-band and the two HST bands, to construct 
spectral energy distributions (SEDs).  We then fit these SEDs to template 
spectra normalized by the near infrared K-band light in a procedure described 
in Brinchmann \& Ellis (2000).   The basic fitting procedure consists
of comparing model SEDs through a $\chi^{2}$ minimization to the best 
fit spectral template from Bruzual \& Charlot (2003) models 
constructed from exponentially declining star formation
rates, Salpeter IMFs, and solar metallicity stellar populations. From this, 
the stellar M/L ratio is determined, and using the K-band
flux, a stellar mass is derived.  Typical uncertainties in this
method are a factor of two (Brinchmann \& Ellis 2000). 

\section{Results} 

\subsection{B-band and K-band Tully-Fisher Relationships}

The TF relationship has been studied in disk 
galaxies out to redshifts $z > 0.5$ by e.g.,
Vogt et al. (1996, 1997, 2004), Ziegler et al. (2002) and Bohm et al. 
(2003). These investigations have found between 0.4 and 1 magnitude of
luminosity evolution in disks at $z > 0.5$ compared to those at $z \sim 0$.  
This luminosity evolution is derived by assuming that the slope of the TF
relationship at high redshift is the same as it is at $z \sim 0$.

\begin{figure}
\vskip -0.1cm
\plotfiddle{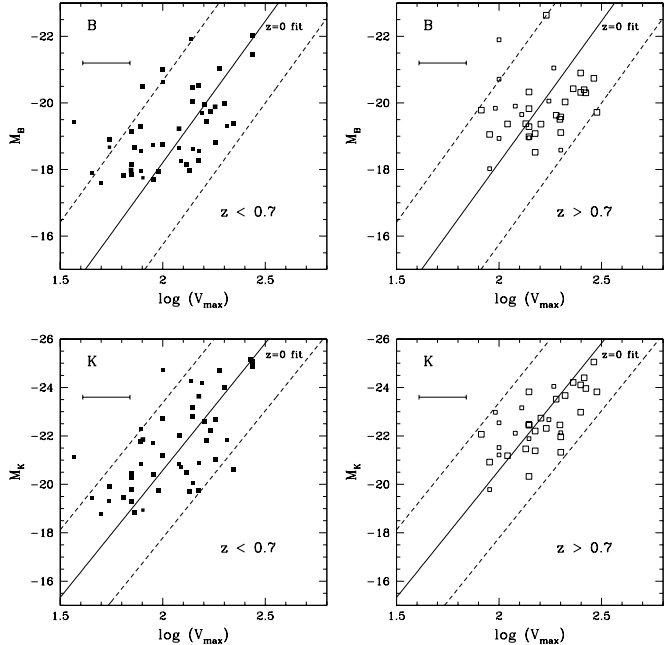}{3in}{0}{44}{44}{-130}{-60}
\caption{Rest-frame B-band and K-band TF relationships for our
sample of disks.  The panels are divided into two redshift
bins, higher and lower than $z = 0.7$.  The solid and dashed lines are 
$z \sim 0$ TF relations found by Verheijen (2001) and their $\pm$ 3 $\sigma$
scatter.  The average error is shown with large/small points having errors 
lower/larger than the average.}
\vskip -0.6cm
\end{figure}

After fitting the TF using using a least-squares method, we find a 
modest luminosity evolution for our $z < 0.7$ disks in comparison to 
the Verheijen (2001) nearby disk sample (Figure~1). By holding 
the $z \sim 0$ slope 
constant for the B-band TF relations, we find a brightening in the zero 
point of $\sim 0.7\pm1.0$ magnitudes in the B-band for disks at $z < 0.7$.  At
$z > 0.7$ we find that the zero point is fainter, by $0.1\pm1.1$.
Neither or these zero point changes are statistically significant,  
similar to the results of Vogt et al. (1997).  The B-band
Tully-Fisher relationship is however degenerate such that we cannot seperate
luminosity evolution from stellar mass growth.  We find the same lack of 
evolution in the K-band Tully-Fisher, with a 
brightening of 0.5$\pm1.2$ magnitudes for
systems at $z < 0.7$, and a fading of 0.1$\pm$1.1 magnitudes for
systems at $z > 0.7$.   The scatter in the Tully-Fisher relationship
also does not appear to change significantly between the B-band or K-band.

\subsection{The Stellar Mass Tully-Fisher Relationship}

By comparing the mass of a dark halo with respect to the stellar mass of its
disk, we can directly measure how these two components evolve
with redshift.  This comparison is to first order 
independent of effects from dust, 
star formation histories, and galaxy sizes and thus offers more physical
insight
than the Tully-Fisher relation.  The classical TF relationship 
between the luminosity of a disk and its maximum velocity scales such
that $L \alpha V^{3.5}$.  This coupling becomes even steeper when
we consider the relationship between stellar mass and velocity in
nearby disks, 
$M \sim V^{4.5}$, in the so-called stellar mass Tully-Fisher relationship
(Bell \& de Jong 2001).
Ideally we would want to measure the gas content of high 
redshift disks to determine their total baryonic mass content.  We can however
use our computed stellar masses to investigate evolution of stellar mass
with virial mass (Bell \& de Jong 2001).

\begin{figure}
\vskip -2.2cm
\plotfiddle{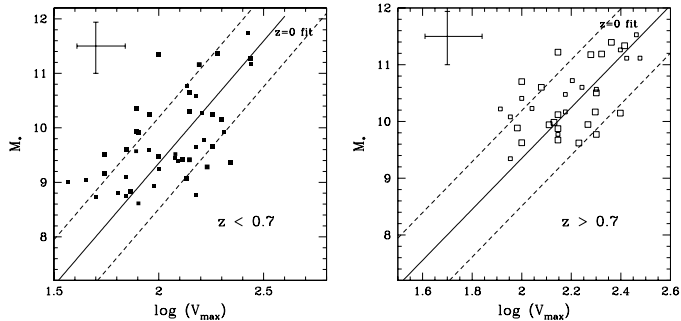}{0.5cm}{0}{45}{45}{-130}{-270}
\vskip 5.3cm
\caption{The stellar mass  Tully-Fisher relationship (M$_{*}$ vs. 
V$_{\rm max}$).  The solid and dashed lines are the $z = 0$ relationship
found by Bell \& de Jong (2001) for nearly disks and its $\pm$ 3 $\sigma$
scatter. The error bar is the
average error with large points having errors lower
than this average, with smaller points having errors larger than the average. }
\end{figure}

We investigate evolution in the stellar-mass TF by 
determining the relationship between V$_{\rm max}$ and M$_{*}$ in the
two redshift bins separated by $z = 0.7$ (Figure~2).  
Each panel shows a solid line which is the relationship 
between V$_{\rm max}$ and M$_{*}$ at $z \sim 0$ as measured by
Bell \& de Jong (2001).
Just as for the B-band and K-band TF relationships there is no significant 
evolution in the stellar mass TF relationship zero
point at either redshift range. After fitting M$_{*}$ with
V$_{\rm max}$ we find essentially the same zero point as the relationship at
$z \sim 0$.   We also find that the scatter in the stellar mass Tully-Fisher
relation appears to be nearly the same as in the K-band Tully-Fisher.

\subsection{Direct Comparison of Stellar and Dark Masses}

We can go beyond the stellar mass Tully-Fisher relationship by 
directly comparing the stellar mass of a halo to its
virial mass.  Figure~3 shows this 
relationship. The fitted relation between the virial and stellar mass 
does not change significantly between low and high redshifts.  There is
also no obvious trend in the scatter from this relationship at either
high or low redshift, although the most depleted systems in
stellar mass tend to have high virial masses, although we cannot yet
say if this is a significant effect or just the result of small
number statistics, or underestimated errors.  Semi-analytic model predictions 
for the ratio of stellar and virial masses is also shown in Figure~3.

\section{Summary}

We demonstrate with ground based and HST data that it is possible to derive 
virial masses from long-slit spectroscopy and high resolution 
imaging,  and stellar masses from
broad-band optical and near infrared photometry.  Based on an analysis
of 83 disk galaxies with both stellar and viral mass estimates,
we are able to trace the evolution of the stellar and dark mass components
of disks out to when they are morphologically identifiable, at $z \sim 1$,
until today. We find the following:

\begin{figure}
\vskip -1.5cm
\plotfiddle{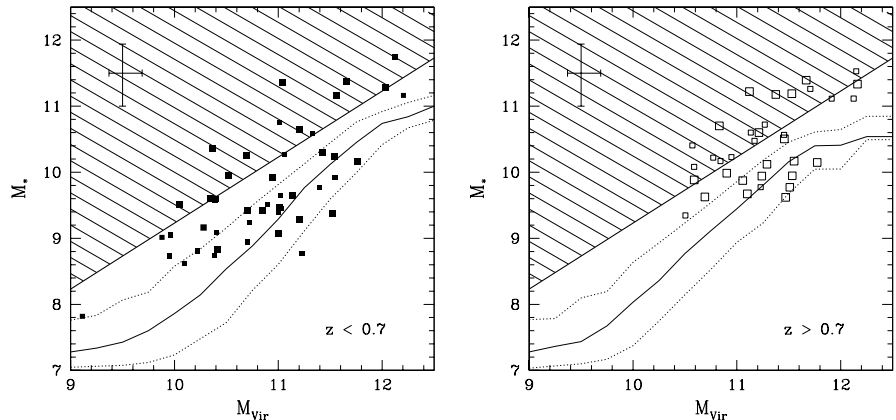}{0.8cm}{0}{60}{60}{-190}{-320}
\vskip 5.3cm
\caption{The relationship between stellar mass (M$_{*}$) and virial mass 
(M$_{\rm vir}$) at redshifts less than and greater than $z = 0.7$. 
The thick solid line is the baryonic fraction
limit and the shaded upper region is where 
M$_{*}$/M$_{\rm vir}$ is larger than the baryonic mass fraction.  The thin 
solid line is the relationship between stellar and virial 
masses from the semi-analytic models of Benson et al. (2002) at $z = 0$
for the $z < 0.7$ sample and at $z = 1.0$ for the $z > 0.7$ sample.
The dashed lines display the 80\% range of where galaxies in
these simulation are found.}
\end{figure}

I. Massive disk galaxies exist out to $z \sim 1$ with virial and stellar
masses as high as 10$^{12}$ \solm, roughly as large as the most massive disks
in the nearby universe.  At least some disk galaxies
are extremely mature even at $z \sim 1$, and appear to be forming
at $z > 1.5$ (Conselice et al. 2003).

II. Similar to other studies, we find no significant evolution in 
the rest-frame B-band or K-band
Tully-Fisher relation out to $z \sim 1.2$.  Both relations 
also contain a similar amount of scatter.

III.  The stellar mass Tully-Fisher relation out to $z \sim 1.2$
is largely consistent with that found for nearby disks.  
We also find no significant evolution in our sample after comparing systems
at redshifts greater than and less than $z = 0.7$. 

IV. We find that the comparison between stellar and virial masses
remains relatively similar from $z \sim 0$ to $z \sim 1.2$, with
no significant change in slope or zero point.

All of these results suggest that at least the brightest disks, which
are morphologically identifiable as disks, are well formed by $z \sim 1.2$.  
This implies
that the major epoch of disk formation, and the establishment of both 
stellar and halo masses, occurs well before this time.

\end{document}